\documentstyle[12pt]{article}
\def\be{\begin{equation}}
\def\ee{\end{equation}}
\def\ket#1{\mbox{$|{#1}\rangle$}}
\begin{document}
\title{THE IRREDUCIBLE STRING AND AN INFINITY OF ADDITIONAL
  CONSTANTS OF MOTION IN A
  DEPOSITION-EVAPORATION MODEL ON A LINE}
\author{M. K. Hari Menon$^1$  and
        Deepak Dhar$^2$ \\
  Tata Institute of Fundamental Research,\\
  Homi Bhabha Road,\\ Bombay-400 005, India.}
\maketitle
\footnotetext[1]{Email:hari@theory.tifr.res.in}
\footnotetext[2]{Email:ddhar@theory.tifr.res.in}
\newpage
\begin{abstract}
We study a model of stochastic deposition-evaporation with
recombination, of three species of dimers on a line.  This model is a
generalization of the model recently introduced by Barma {\it et. al.}
(1993 {\it Phys. Rev. Lett.} {\bf 70} 1033) to $q\ge 3$ states per
site. It has an infinite number of constants of motion, in addition to
the infinity of conservation laws of the original model which are
encoded as the conservation of the irreducible string. We determine
the number of dynamically disconnected sectors and their sizes in this
model exactly.  Using the additional symmetry we construct a class of
exact eigenvectors of the stochastic matrix.  The autocorrelation
function decays with different powers of $t$ in different sectors.  We
find that the spatial correlation function has an algebraic decay with
exponent $3/2$, in the sector corresponding to the initial state in which
all sites are in the same state.  The dynamical exponent is nontrivial in
this sector, and we estimate it numerically by exact diagonalization
of the stochastic matrix for small sizes.  We find that in this case
$z=2.39\pm0.05$.
\end{abstract}

PACS Numbers: 02.50+s, 75.10J, 82.20.M, 05.50+q
\newpage

\section{Introduction}
Recently a very interesting stochastic model with deposition and
evaporation processes has been introduced by Barma {\it et. al.}
\cite{mbg1,mbg2}.  In this model one deposits atoms on to $k$ adjacent
vacant sites of a $d$-dimensional lattice and evaporates atoms from
any $k$ adjacent occupied sites, with specific rates for deposition
and evaporation. The cases $k=1$ and $k=2$ are exactly solvable,
the former being trivial, and the latter being equivalent to
ferromagnetic Heisenberg spin-$\frac{1}{2}$ chain.
For $k\ge 3$, on a linear chain of length $L$, the
phase space of this $k$-mer model consisting of the $2^L$ possible
configurations is found to break up into an exponentially large number
of dynamically disconnected sectors \cite{mbg1,mbg2}. This may be
understood as being due to the existence of an infinite number of
independent conserved quantities in this model \cite{ddmb1}.  These
conservation laws also give rise to a wide range of relaxation
behavior \cite{ddmb2}. For example the density-density autocorrelation
function shows different power law decays in different sectors
($t^{-1/4},t^{-1/2},t^{-0.59}$) and even a stretched exponential decay
in some sectors.

For $k\ge 3$, though there exist an infinite number of constants of
motion, these still have not enabled us to get a full solution of the
model so far. In this case the number of sectors and their sizes in the
steady state has been calculated exactly, but the dynamics is
understood only qualitatively. The quantum Hamiltonian corresponding
to the stochastic matrix has $k$-body interaction terms which makes
the problem difficult to tackle analytically.  Thus it is worthwhile
studying models which show the same qualitative features and are at
the same time more tractable analytically.

With this motivation we define a variant of the $k$-mer
deposition-evaporation model with $q$ states per site.  This model for
$q=2$ corresponds to the dimer deposition evaporation model studied by
Barma {\it et.al}. For $q\ge2$ this model shares many qualitative
features with the trimer model, like the existence of an infinity of
conservation laws encoded as the conservation law of the ``irreducible
string'' \cite{ddmb1}, and different power law decay of the
autocorrelation function in different sectors. However our model has
the two advantages: it has only $2$-body interaction and it has an
additional discrete symmetry group of very large order. These
properties make this model more tractable analytically than the
$k$-mer model.  In fact, one can get many exact eigenvectors of the
stochastic matrix in our model very easily.  However, we have not
succeeded so far in solving this model completely. For simplicity
we study only $q=3$ case in this paper, for higher $q$  results
will be qualitatively the same.

This paper is organized as follows. In section $2$ we define our model
and write the stochastic evolution operator in the form of the
Hamiltonian of a quantum many-body system of $3$ species of particles.
In section $3$ we show that the vector space consisting of $3^L$
configurations breaks up into an exponentially large number of
dynamically disconnected sectors, and also calculate their sizes. In
section $4$ we describe the additional symmetries of this model and
study the resulting further decomposition of the phase space formed by
all the configurations belonging to a sector into subspaces.  We
construct a class of exact eigenvectors in the next section. In
section $6$ we study the decay of autocorrelation function in various
sectors.  As in the case of trimer model \cite{ddmb2}, we find sector
dependent decay.  The spatial correlation function in the steady state
for the sector corresponding to an initial condition of all sites in
the same state is calculated in section $7$. It has power law decay
with exponent $3/2$.  A numerical diagonalization study of the
stochastic matrix to find the dynamical exponent in this sector is
presented in section $8$.

\section{Definition of the model}
Our model is defined as follows: At each site of a one dimensional
lattice, there is a spin variable $q_i$ which can be in any of the
three states $a,b$ or $c$.  These spin variables undergo a stochastic
time evolution given by the following rule: any pair of adjacent
spins, which are in the same state, can flip together to any of the
other two states with some specified rates for the transitions.  For
example an $aa$ pair can flip to become either a $bb$ pair or a $cc$
pair. This process can be thought of as the evaporation of a $aa$
dimer and immediate deposition of a $bb$ or $cc$ dimer at that place.
We shall call this model the dimer deposition-evaporation model (DDE model).
This is a special case of the general reaction-diffusion process
($a+b\rightleftharpoons c+d$) recently studied by Dahmen \cite{dahmen}.
We consider only the case of equal rates for all the transitions in this
paper.

Any configuration $C$ on the lattice can be represented by the basis
vector $\ket{q_1,q_2,\ldots,q_L}$, in a
vector space of dimension $3^L$. If $P(C,t)$ denote the probability of
the configuration $C$ at time $t$, then the master equation describing
the evolution of these probabilities can be written as
\be
\frac {\partial}{\partial t}
\ket{P(t)} =\hat W \ket{P(t)} \\
\ee
where $\ket{P(t)}=\sum_CP(C,t)\ket{C}$ and $\hat W$ is the stochastic
matrix.

We can write $\hat W$ in the form of a quantum Hamiltonian as
follows. Consider a quantum mechanical system of 3 species of
particles on a one dimensional lattice of length $L$ with periodic
boundary condition. Each species is characterized by a color $q$ which
can be either $a,b$ or $c$. Particles of the same colour experience a
hard core interaction so that not more than one particle of the same
color can occupy a given site. Therefor every site can be either empty
or occupied by up to $3$ particles of different colors.  Thus there
are $8$ possible states on every site.  Let $\hat q_i$ and $\hat
q_i^+$ denote the Pauli operators which annihilate and create a
particle of color $q$ at the site $i$. Now consider the Hamiltonian
\be
\hat H = \sum_{i=1}^L \hat A_{i,i+1} + \hat B_{i,i+1} + \hat C_{i,i+1}
\label{hami}
\ee
where
\[
\begin{array}{ccc}
\hat A_{i,i+1} & = &(\hat b_i^+ \hat b_{i+1}^+  + \hat c_i^+ \hat c_{i+1}^+
- 2\hat a_i^+ \hat a_{i+1}^+) \hat a_i \hat a_{i+1} \\
\hat B_{i,i+1} & = &(\hat a_i^+ \hat a_{i+1}^+ + \hat c_i^+ \hat c_{i+1}^+
- 2\hat b_i^+ \hat b_{i+1}^+) \hat b_i \hat b_{i+1} \\
\hat C_{i,i+1} & = &(\hat a_i^+ \hat a_{i+1}^+  + \hat b_i^+ \hat b_{i+1}^+
- 2\hat c_i^+ \hat c_{i+1}^+) \hat c_i \hat c_{i+1}
\end{array}
\]
where the operator $\hat A_{i,i+1}$  acting on any state results a
nonnull vector only when $q_i=q_{i+1}=a$ and is given by,
\be
\hat A_{i,i+1} \ket{\ldots,a,a,\ldots}=
\ket{\ldots,b,b,\ldots} + \ket{\ldots,c,c,\ldots} - 2\ket{\ldots,a,a,\ldots}
\label{action}
\ee
The operators $\hat B_{i,i+1}$ and $\hat C_{i,i+1}$ have similar
actions. The operator
\be
\hat n_i = \hat a_i^+\hat a_i + \hat b_i^+\hat b_i +
           \hat c_i^+\hat c_i
\ee
counts the number of particles at site $i$.
Clearly
\be
[\hat n_i, \hat H]=0,
\ee
and the number of particles is conserved at each site.
If we restrict ourselves to the subspace of the full
Hilbert space in which there is only one particle at every site, {\it ie.}
$\hat n_i=1$ for all $i$,
corresponding to every configuration $\ket{q_1, q_2, \ldots, q_L}$ in
this subspace there is a unique configuration specified by the same
L-string in the dimer problem.  The action of $\hat H$ on
$\ket{\{q_i\}}$ is the same as that of $\hat W$.
Thus $\hat W$ is represented by the quantum Hamiltonian
given by (\ref{hami}) acting on the subspace with $\hat n_i=1$
for all $i$.

Make the particle-hole transformation ($\hat q_i \rightarrow \hat q_i^+$)
on all odd numbered sites for all $q$. We define the  current operator
by
\be
\hat J_{i,i+1} = \hat a_i^+ \hat a_{i+1} + \hat b_i^+ \hat b_{i+1}
                  +\hat c_i^+ \hat c_{i+1}.
\ee
Then the Hamiltonian up to the  addition of a constant can be written as
\be
\hat H = \sum_i \hat J_{2i,2i+1} \hat J_{2i,2i+1}^+ + \hat J_{2i,2i-1}
         \hat J_{2i,2i-1}^+ + 3 \sum_{i,\alpha} n_{i,\alpha}
         n_{i+1,\alpha}.
\ee
The dimer model now correspond to $\hat H$ operating on a sector
where $\hat n_i=1$ for all even numbered sites and $\hat n_i=2$ for all
odd numbered sites. The corresponding stochastic process is one
in which there is a constant rate of exchange of any two particles
at nearest neighbour sites.
\section{The Sector Decomposition of the Vector Space}
As the original trimer model, this model has an infinite number of
independent constants of motion. These are described most simply
in terms of the ``irreducible  string''(IS) \cite{ddmb1}.
We  consider free boundary conditions for convenience.
Each configuration can be
represented as a string of $L$ characters, each character
being one of $a,b$ and $c$. From the string
corresponding to a given configuration C, delete all adjacent pairs of
$aa$, $bb$ and $cc$.  Each such deletion decreases the length of the
string by $2$.  Repeat this operation on the resulting string until
no more deletions are possible.  This defines the  IS corresponding
to C.

It is straightforward to extend the arguments of \cite{ddmb1}
to our model, and show that the IS is a constant of motion. In
addition, any two configurations having the  same IS can be
reached from each other. Thus the IS can be  used to label the
different sectors into which the vector space breaks up.

Consider the set of all configurations in which no adjacent pair of
sites are in the same state. These configurations will not
evolve under the given dynamics and are said to be  fully
jammed. For these  configurations the length $l$ of the irreducible
string  equals $L$.
The total number of such configurations is
easily seen to be equal to $3\times{2^{L-1}}$.  Now consider sectors which
are not totally jammed ($l<L$). The number of sectors labelled by
irreducible strings of length $l$ will be the number of distinct
irreducible strings of length $l$ which is $3\times{2^{l-1}}$.
The total number of sectors can be easily seen to be
$2^{L+1}-1$.

Now we proceed to calculate the sizes of these sectors, {\it i.e.} the
number of distinct configurations which belong to a particular sector.
The size of a fully jammed sector is clearly $1$. This is because a
totally jammed configuration can not evolve and hence is the only
member of that sector. Let $D(IS,L)$ be the size of the sector
labelled by irreducible string $IS$ on a lattice of length $L$. We
define the generating function
\be
G(IS,z) = \sum_{L=l}^{\infty} D(IS,L)z^L
\ee
where the summation over $L$ extends from $l$, the length of $IS$ to
infinity.  Of special interest is the sector corresponding to IS of
length zero.  This we will call the null sector and denote its
irreducible string by $\phi$.  To compute $G(\phi,z)$, we introduce a
formal series which is a sum of all strings of arbitrary length, which
are made up of three letters $a$, $b$ and $c$, and reducible to $\phi$.
\be
\tilde G(\phi)=\phi+aa+bb+cc+aaaa+bbbb+cccc+abba+\ldots
\ee
A string $S$ is said to be {\it decomposable}, if the irreducible
string $IS(S)=\phi$ and it can be written as $S_1.S_2$
such that $IS(S_1)=IS(S_2)=\phi$.  If
$\tilde G_I(\phi)$ denote the sum of all {\ indecomposable} strings
which are reducible to $\phi$ then $\tilde G(\phi)$ satisfies the equation
\be
\tilde G(\phi)=\frac{\phi}{1-\tilde G_I(\phi)} \label{eq:G1}
\ee
Further if $\tilde G_I^q$
denotes the the sum of all indecomposable strings which are reducible
to $\phi$ and starting with the letter $q$ then
\be
\tilde G_I(\phi)=\tilde G_I^a+\tilde G_I^b+\tilde G_I^c
\label{eq:G2}
\ee
$\tilde G_I^q$ is given by
\be
\tilde G_I^q=q~.~\frac{1}{1-(\tilde G_I^a + \tilde G_I^b + \tilde G_I^c-
\tilde G_I^q)}~.~q
\label{eq:G2.1}
\ee
To get the generating function we replace all occurences
of $a,b,c$ by $z$ in equation (\ref{eq:G2.1}).
Then $\tilde G_I^a,\tilde G_I^b$ and
$\tilde G_I^c$ all reduce to the same power series in $z$,
call it $g(z)$, given by
\be
g(z) = \frac{z^2}{1-2g(z)}
\ee
This equation determine  $g(z)$ as an explicit function of $z$
given by
\be
g(z) = (1 - \sqrt{1-8z^2})/4.
\label{eq:g}
\ee
Using equations (\ref{eq:G1}) and (\ref{eq:G2}), we get
\be
G(\phi,z)=\frac{1}{1-3g(z)}
\ee
The growth of $D(\phi,L)$ for large $L$ is determined by the
singularities of $g(z)$ nearest to the origin. This happens at
$z_c=\pm 1/\sqrt{8}$.
And since $g$ has a square root singularity near $z_c$,
it is easily  verified that
\be
D(\phi,L)\sim 8^{L/2}~L^{-3/2}
\ee
for large $L$.

One can easily generalize this procedure to find the size of the
sector when the irreducible string is of finite length $l$.
Let $IS=\alpha_1\alpha_2\ldots,\alpha_l$, where
$\alpha_i=a,b$ or $c$. Then $\tilde G(IS)$ (defined similarly as
$\tilde G(\phi)$) is given by
\be
\tilde G(IS)=\frac{1}{1-\tilde H(\alpha_1)}.\alpha_1.\frac{1}{1-\tilde
  H(\alpha_2)}.\alpha_2.\ldots.\frac{1}{1-\tilde
  H(\alpha_l)}.\alpha_l.\tilde G(\phi)
\ee
where $\tilde H(\alpha)=\tilde G_I^a+\tilde G_I^b+\tilde G_I^c-
\tilde G_I^{\alpha}$.
Replacing $a,b,c$ by $z$ as before we get
\be
G(IS,z)={[\frac{z}{1-2g(z)}]}^l\frac{1}{1-3g(z)}
\label{eq:g1}
\ee
where $g(z)$ is given by  equation (\ref{eq:g})
and $l$ is the length of the irreducible
string $IS$. Then the size of the sector characterised by the
irreducible string IS is the coefficient of $z^L$ in (\ref{eq:g1}),
and hence
\be
D(IS,L)=\frac{1}{2\pi i}\oint_{c} dz\frac{G(IS,z)}{z^{(L+1)}}
\ee
where integration is over a small circle encircling the origin.
{}From this it is straightforward to find the asymptotic behavior of
$D(IS,L)$ by doing the  integral using the saddle point approximation.
Let $\lambda=\lim_{L \rightarrow \infty} l/L$.
Then
\be
D(IS,L)\sim{[k(\lambda)]}^L,
\label{sec_size1}
\ee
where
\be
k(\lambda)=(1-\lambda)^{-(1-\lambda)/2}~(1+\lambda)^{-(1+\lambda)/2}
{}~2^{(3-\lambda)/2}.
\label{sec_size2}
\ee
Thus, in general the size of a sector increases exponentially with $L$
and the growth constant depends on the density of the irreducible string.
{}From this one can recover the results for the null and totally jammed
sectors by noting that $k(0)=\sqrt{8}$ and $k(1)=1$.

\section{Additional Symmetries}
In the previous section we discussed the decomposition of the $3^L$
dimensional configuration space of the model into $2^{L+1}-1$ mutually
disjoined sectors. This sector decomposition described in terms of the
irreducible string $IS$ is maximal in the sense that it is possible to
go from any configuration to any other configuration in the same
sector. However, this only helps us to block diagonalize $\hat W$. In
order to do better, we have to find additional operators that commute
with $\hat W$.  These operators are not completely diagonal in the
configuration basis.  We now describe how to construct such operators
using the additional symmetries of our model. These symmetries are not
present in the original $k$-mer model.  They are also not present in
our model if all the transition rates are not equal. Using these
symmetries we are able to reduce the task of diagonalizing the
stochastic matrix with in a sector, by further block diagonalizing it.

The  deposition-evaporation model can be recast in the form of
a generalized interface growth model, similar to the
KPZ model,  where the variable  at each site is
a $2\times2$ matrix instead of a scalar. This is seen as follows:
we consider three non commuting matrices $A(a),A(b)$ and $A(c)$
such that $A^2(a)=A^2(b)=A^2(c)=I_D$, the identity
matrix. A simple choice which satisfies these conditions is
the unimodular matrices
\be
\begin{array}{rclrclrcl}
A(a)&=&\left(\begin{array}{cc}1&0\\1&-1\end{array}\right),&A(b)&=&\left(
  \begin{array}{cc}1&1\\0&-1\end{array}\right),
A(c)&=&\left( \begin{array}{cc}1&0\\0&-1\end{array}\right)
\end{array}
\ee
Alternately, we could choose them to be $SU(2)$ unitary matrices.
For a configuration
$\ket{C}=\ket{q_1,q_2,\ldots,q_L}$, we associate a $2\times2$ matrix
$I_i(C)$ to every site $i$ given by
\be
I_i(C) = A(q_i)\ldots A(q_2)A(q_1).
\ee
It is easy to see that the stochastic evolution
of the dimer model can be cast in terms of matrices $I$'s
by the following local evolution rule: if $I_{i-1}=I_{i+1}$, then
$I_i$ is reset randomly to any one of the three
values $A(q)I_{i-1}$ at a constant rate say $1$. The
conservation  law of the $IS$ in this representation becomes the
simple statement that $I_L$ is unchanged.

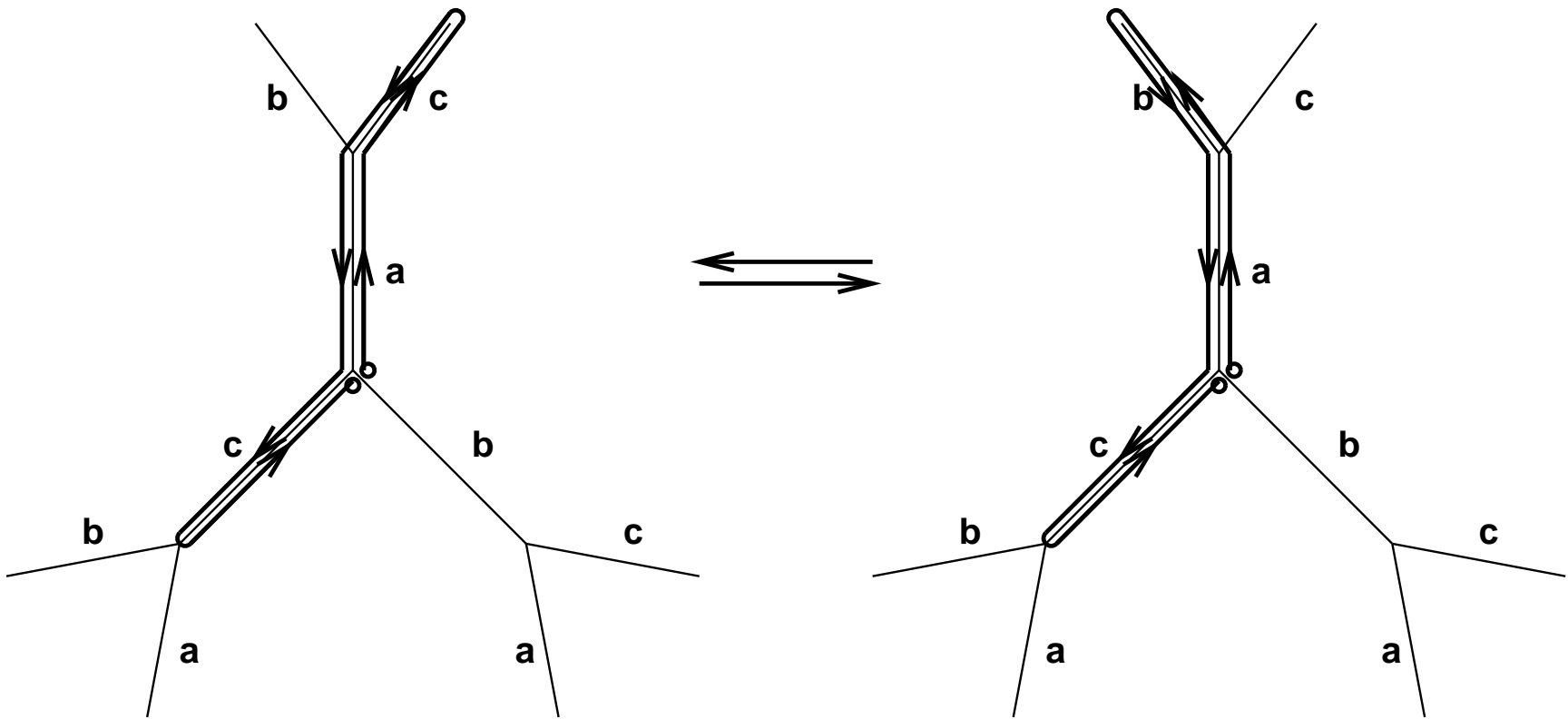
\begin{figure}
\caption{Dimer deposition-evaporation dynamics as a
dynamics of polymer chain on the Bethe lattice. The transition between the
states $\ket{accacc}$ and $\ket{abbacc}$ is shown.}
\label{b_l_rep}
\end{figure}

A configuration of dimers is then equivalently characterized by
specifying the matrix variables $I_i$ at each site $i$. This matrix
formulation brings out clearly the existence of a discrete group of
infinite order in the model.  If we work with free boundary conditions
the set of allowed values of the of the matrices $I$'s can be put in
one to one correspondence with the sites of a $3$-coordinated Bethe
lattice.  Identify one site of the Bethe lattice as the origin and
associate the identity matrix to this site. We color the bonds of this
Bethe lattice by three colors $a,b,c$ such that any three bonds
meeting at a site are all of different colors. The matrix $I_{\alpha}$
corresponding to a site $\alpha$ of the Bethe lattice is given by the
ordered product of $A(q)$s $(q=a,b,c)$ along the unique path from the
origin to $\alpha$. A sequence $\{q_i\}$ , $i=1$ to $L$ specifies a
configuration of the DDE model, and also a unique $L$-step path on the
Bethe lattice which starts at the origin.  As an example, the path
corresponding to the configuration $accacc$ is shown in figure
\ref{b_l_rep}.  Configurations in the null sector correspond to paths
that return to the origin. Any pair of adjacent spins that are in the
same state in the dimer configuration will corresponds to an immediate
retraversal of a step on the Bethe lattice.  The stochastic evolution
in the DDE model leads to a stochastic evolution of this $L$-step
polymer chain on the Bethe lattice.  An elementary movement of this
polymer chain dynamics is illustrated in figure \ref{b_l_rep}, where
the transition from the configuration $acccacc$ to $abbacc$ is
shown.

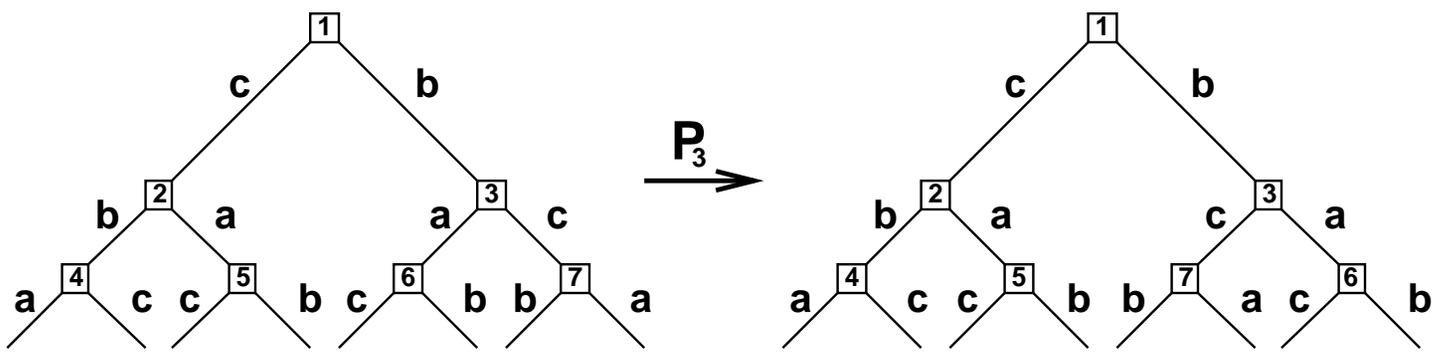
\begin{figure}
\caption{The operation of interchange of two branches that start
from the site $3$. This is equivalent to recoloring of the bonds,
and is a symmetry of $\hat W$.}
\label{br_flip}
\end{figure}

The operation of interchange of one branch of the Bethe
lattice with another is a symmetry of $\hat W$ since this only
corresponds to a recoloring of the bonds. This is illustrated in
figure \ref{br_flip}, where we have considered only one branch
starting from the origin.  At every site $\alpha$ on the Bethe lattice
define an operator $\hat P_\alpha$ which interchanges the two subtrees
starting from that site and going away from the origin.  $\hat
P_\alpha$ changes a configuration
$C\equiv\{I_i\}$ to $C'\equiv\{I_i'\}$ where
$I_i'\rightarrow f(I_i)$, where $f$ is a matrix valued function of its
argument.

Clearly $[\hat W,\hat P_\alpha]=0$.  We say that site $\beta$ is
descendant of site $\alpha$ if the unique path connecting $\beta$ to
the origin goes through $\alpha$. From the figure it can be seen that
$[\hat P_\alpha,\hat P_\beta]=0$ if neither $\alpha$ nor $\beta$
is a descendant of the
other. If $\beta$ is a descendant of $\alpha$ then $\hat
P_\alpha\hat P_\beta = \hat P_\gamma \hat P_\alpha$, where $\hat
P_\gamma$ is the operator at site $\gamma$ which is obtained from site
$\beta$ by the action of $\hat P_\alpha$.  In figure \ref{br_flip} for
example, $[\hat P_2,\hat P_3]=0$.  But $[\hat P_1, \hat P_2]\ne 0$
since $2$ is a descendant of $1$. Instead they satisfy
the relation $\hat P_1 \hat P_2 = \hat P_3 \hat P_1$. This symmetry
comes from the geometrical symmetry of the underlying Bethe lattice.
This in turn is related to the internal color symmetry in the model,
and we will call it recoloring symmetry C.

In addition to these one has a permutation symmetry of order $3$
($S_3$) about the origin which corresponds to interchanging the $3$
color labels globally. The set of operators $\{\hat P_\alpha\}$
together with the operators corresponding to $S_3$ will constitute an
infinite set of conserved quantities, which are however not all
commuting with each other.

One interesting consequence of the recoloring symmetry is that the spectrum
of $\hat W$ in a sector depends only on the length $l$
of the irreducible string and not on its details. This
can be seen as follows: For a general sector, the endpoints of all the
paths on the Bethe lattice will be fixed.  The shortest path
connecting these two points corresponds to the irreducible string. All
$\hat P_{\alpha}$s such that site $\alpha$ does not belong to this
path leaves the irreducible string and hence the sector it labels
invariant. On the other hand one can go from one irreducible string to
any another one having the same length by a series of branch flip
operations using the $\hat P_{\alpha}$ operators which are on the
shortest path.  Since each of these operations are symmetry operations
of $\hat W$, {\it i.e.}  $[\hat P_{\alpha},\hat W]=0$ for all
$\alpha$, the spectrum will be identical for all these sectors.  This
property is not present in the original $k$-mer model.

We can make use of these symmetries for further block diagonalisation
of $\hat W$ with in a sector.  This is achieved by breaking the state
space spanned by all the configurations in a particular sector into
subspaces.  We illustrate this with a few examples.

For convenience we will consider the case of null sector.  First
consider a lattice of length $2$. There are only three configurations
in this sector which are $\ket{aa},\ket{bb}$ and $\ket{cc}$.  In this
case we can get all the eigenstates of the Hamiltonian using
$S_3$. Consider the states
\be
\begin{array}{lcl}
\ket{\psi_1}&=&\ket{aa}+\ket{bb}+\ket{cc} \\
\ket{\psi_2}&=&\ket{aa}+\omega \ket{bb}+\omega^2 \ket{cc} \\
\ket{\psi_3}&=&\ket{aa}+\omega^2 \ket{bb}+\omega \ket{cc}.
\end{array}
\ee
where $\omega$ is the cube root of unity.
These are the eigenstates of the operator corresponding to cyclic
permutation of $a$, $b$ and $c$ ($\hat P_{abc}$) with eigenvalues
$1,\omega$ and $\omega^2$ respectively.
Since $[\hat P_{abc},\hat
W]=0$ these are also the eigenstates of $\hat W$.
The eigenvalues of $\hat W$ corresponding to these
eigenvectors are $0,-3,-3$.
Thus in this case we were able to diagonalize $\hat W$ completely.
But this is a rather trivial example.

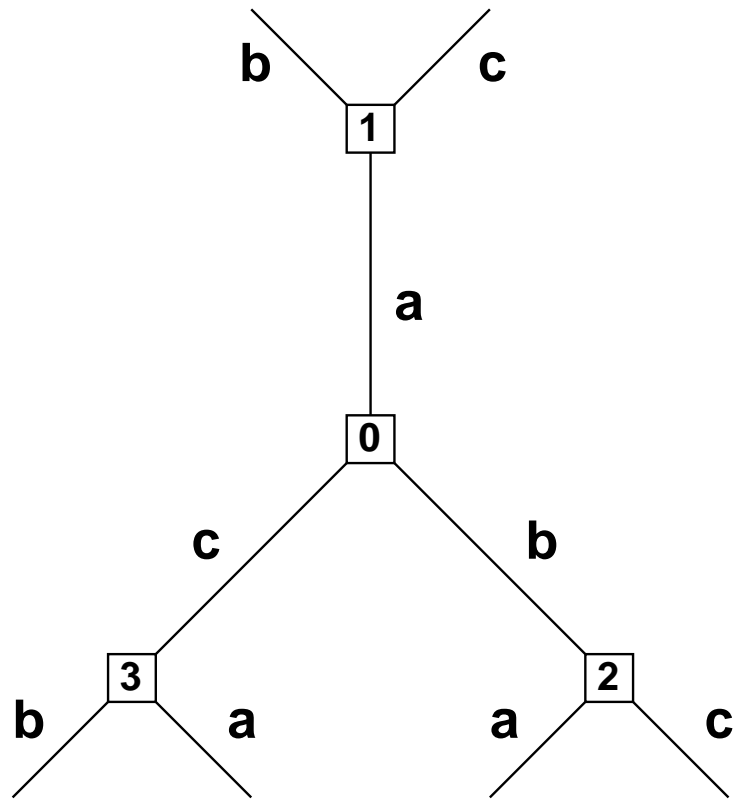
\begin{figure}
\caption{Construction of exact eigenvectors using recoloring
symmetry. For $L=4$, three exact eigenvectors can be constructed by
antisymmetrizing about the sites $1,2$ and $3$.}
\label{bl_l4}
\end{figure}

Next we consider the case of $L=4$. There
are $15$ states in this sector. First we consider the symmetry
operations about the sites $1$, $2$
and $3$ as shown in figure \ref{bl_l4}.
Consider the states
\be
\begin{array}{lcl}
\ket{\psi_1}&=&\frac{1}{\sqrt{2}}(\ket{abba}-\ket{acca})\\
\ket{\psi_2}&=&\frac{1}{\sqrt{2}}(\ket{bccb}-\ket{baab})\\
\ket{\psi_3}&=&\frac{1}{\sqrt{2}}(\ket{caac}-\ket{cbbc}).\\
\end{array}
\label{psi}
\ee
These are the non-degenerate
eigenstates with eigenvalue $-1$ of $\hat P_1$, $\hat P_2$
and $\hat P_3$ respectively. It can be verified very easily that these are
also eigenstates of $\hat W$. This gives us a decomposition of the
$15$ dimensional vector space into subspaces of
dimensions $1,1,1$ and $12$. Using the cyclic permutation
symmetry at the origin, we can get a further decomposition of
the $12$ dimensional subspace. For this consider the eigenstates of
$\hat P_{abc}$
\be
\begin{array}{lcl}
\ket{\phi_1}&=&\frac{1}{\sqrt{3}}(\ket{aaaa}+\ket{bbbb}+\ket{cccc})\\
\ket{\phi_2}&=&\frac{1}{\sqrt{3}}(\ket{aabb}+\ket{bbcc}+\ket{ccaa})\\
\ket{\phi_3}&=&\frac{1}{\sqrt{3}}(\ket{bbaa}+\ket{ccbb}+\ket{aacc})\\
\ket{\phi_4}&=&\frac{1}{\sqrt{6}}(\ket{abba}+\ket{acca}+\ket{bccb}\\
            & &+~~\ket{baab}+\ket{caac}+\ket{cbbc})\\
            & &                                                    \\
\ket{\chi_1}&=&\frac{1}{\sqrt{3}}
(\ket{aaaa}+\omega\ket{bbbb}+\omega^2\ket{cccc})\\
\ket{\chi_2}&=&\frac{1}{\sqrt{3}}
(\ket{aabb}+\omega\ket{bbcc}+\omega^2\ket{ccaa})\\
\ket{\chi_3}&=&\frac{1}{\sqrt{3}}
(\ket{bbaa}+\omega\ket{ccbb}+\omega^2\ket{aacc})\\
\ket{\chi_4}&=&\frac{1}{\sqrt{6}}
((\ket{abba}+\ket{acca})+\omega(\ket{bccb}\\
            & &+~\ket{baab})+\omega^2(\ket{caac}+\ket{cbbc}))\\
            & &                                                    \\
\ket{\xi_1}&=&\frac{1}{\sqrt{3}}
(\ket{aaaa}+\omega^2\ket{bbbb}+\omega\ket{cccc})\\
\ket{\xi_2}&=&\frac{1}{\sqrt{3}}
(\ket{aabb}+\omega^2\ket{bbcc}+\omega\ket{ccaa})\\
\ket{\xi_3}&=&\frac{1}{\sqrt{3}}
(\ket{bbaa}+\omega^2\ket{ccbb}+\omega\ket{aacc})\\
\ket{\xi_4}&=&\frac{1}{\sqrt{6}}
((\ket{abba}+\ket{acca})+\omega^2(\ket{bccb}\\
            & &+~~\ket{baab})+\omega(\ket{caac}+\ket{cbbc}))\\
\end{array}
\label{phi}
\ee
Since $\hat P_{abc}$ commutes with $\hat W$, $\{\ket{\phi_i}\}$,
$\{\ket{\chi_i}\}$ and $\{\ket{\xi_i}\}$  form $3$ invariant
subspaces under the action of $\hat W$. Thus we have got the
following  decomposition
of the $15$ dimensional vector space spanned by all the configurations in the
null sector in to subspaces of dimensions.
\be
1 + 1 + 1 + 4 + 4 + 4=15
\ee
We have carried out the same procedure for the case of $L=6$.  In this
case the size of the null sector is $87$.  Using the $\hat P_{\alpha}$
operators, this can be broken into 18 subspaces out of which $12$ are of
size $1$, $3$ are of size $7$ and another $3$ are of
size $18$.

A further decomposition of these subspaces can be achieved using
the symmetry of dynamics under reflection. Using both the
color and reflection symmetries, the resulting form of the decomposition
of the vector space  for the case $L=6$ is
\be
87 = 12\times 1 + 3\times 3 + 3\times 4 + 3\times 6
+ 3 \times 12.
\ee
where $m\times n$ means $m$ subspaces of size $n$.

In principle this procedure can be carried out for a lattice of any
length. For a linear chain of length $L$, sites up to distance $L/2$
from the origin on the Bethe lattice are reachable.  The number of
sites that are one step away from the boundary and inside will be
$3\times 2^{(L-4)/2}$ ($L\ge 4$).  As we have seen for $L=4$, the
antisymmetric eigenstates of the $\hat P$ operators on each of these
sites will be an eigenstate of $\hat W$. Thus one can easily get a
large number of eigenstates in this model. By constructing the
antisymmetric eigenstates of the $\hat P$ operators on sites which are
interior to this, one can get invariant subspaces as we have seen in
the case of $L=6$ (the 3 sectors of size 7). So far we have not been
able to get a general result for the total number of subspaces and
their sizes within a sector as a result of this symmetry. But one can
use this symmetry for the numerical diagonalization of the stochastic
matrix for small sizes, where one can restrict oneself to one subspace
(say the symmetric subspace where all the eigenvalues of the $\hat
P$ operators are $1$). This we have done to estimate the dynamical
exponent in the null sector, where we have seen that these additional
symmetries lead to substantial decrease in the size of the matrices to
be diagonalized. These results are discussed in section $8$

\section{Construction of some exact eigenvectors of the
stochastic matrix} In the previous section we have constructed some of
the eigenvectors of the stochastic matrix in the null sector using
the recoloring symmetry of the dynamics of
$L$-step chain on the Bethe lattice. In this section we show
that the same procedure can be used to construct eigenvectors in
sectors which are almost totally jammed, {\it i.e.} the length of the
irreducible string $l=L-2n$, where $n$ is a small integer.

As we explained in the earlier section, the spectrum of $\hat W$ is
the same for all the sectors whose irreducible strings are of same
length.  So we need to consider only one representative sector which
we choose to be the one in which the irreducible string is
$abab\ldots$. Consider the case when $l=L-2$. We use periodic boundary
condition for convenience.  There are two types of configurations in
this sector. In the first type there will be three adjacent sites, say
$i,i+1$ and $i+2$ which are of the same color (either $aaa$ or $bbb$
{\it e.g.}  $\ket{\ldots abbbab\ldots }$). Let these states be denoted
by $\ket{\phi (i)}$. In the second type there will be two adjacent
sites, say $i+1$ and $i+2$ which are having color $c$.  ({\it e.g.}
$\ket{\ldots abccab\ldots}$. Let this type of states be denoted by
$\ket{\chi (i)}$. In both these types site $i+1$ and site $i+2$ have
the same color and thus constitute a reducible part. It can be very
easily seen that under the deposition-evaporation dynamics, the
position of the reducible part makes a nearest neighbour random walk.
Hence the dynamics in this sector is equivalent to a random walk
problem with two states per site ($\ket{\phi(i)}$ and
$\ket{\chi(i)}$). The eigenfunctions $\ket{\psi_k}$ and eigenvalues
$\lambda_k$ of the stochastic matrix can be found by solving this
random walk problem.

Let $\phi_k(i)$ and
$\chi_k(i)$ be defined by $\phi_k(i)=\langle \phi(i)|\psi_k\rangle$ and
$\chi_k(i)=\langle \chi(i)|\psi_k\rangle$. They satisfy the following
eigenvalue equations
\be
\begin{array}{lcl}
\lambda_k \phi_k(i)&=&\phi_k(i-1) + \phi_k (i+1) +
\chi_k(i-1) + \chi_k(i) -4 \phi_k(i) \\
                   & & \\
\lambda_k \chi_k(i)&=&\phi_k(i) + \phi_k(i+1) - 2 \chi_k(i)
\end{array}
\ee
The solution for this eigenvalue equations are given by
\be
\lambda_k=(3-\cos(k)) \pm\sqrt{3+\cos^2(k)}
\ee
and
\be
\begin{array}{lcl} \phi_k(n)&=&e^{ikn} \\
\chi_k(n)&=&\frac{1}{\lambda_k+2}(e^{ikn}(1+e^{ik})) \\
\end{array}
\ee
where $k=2\pi m/L, m=0,1,\ldots,L-1$.

In the case of free boundary conditions some of the localized
eigenvectors
can be constructed directly with out solving this random walk
problem.  Consider the situation where the reducible part is near
one of the boundaries. {\it i.e.} the states $\ket{ab\ldots
abaa}, \ket{ab\ldots abbb}$ and $\ket{ab\ldots abcc}$. It is easy to see
that the state obtained by antisymmetrizing, $\ket{\psi}=\ket{ab\ldots
abaa} -\ket{ab\ldots abcc}$, is an eigenstate with eigenvalue $-1$.
One more eigenstate can be constructed by antisymmetrizing about the
other end.

Now consider the case of $l=L-4$. In this case the reducible part is
of length $4$. A typical configuration in this sector will be
$\ket{\ldots ab\alpha\alpha ab\ldots ab\beta\beta ab\ldots}$.  The
reducible parts $\alpha\alpha$ and $\beta\beta$ perform random walk
under the dynamics. When these random walkers come closer to each
other they can form state like $\ket{\ldots ab\alpha\alpha\alpha\alpha
ab\ldots} $ or an intermediate state like $\ket{\ldots
abc\beta\beta cab\ldots}$. There are two such intermediate states
which corresponds to the two values $\beta=a, b$. The two random
walkers can go into such intermediate states and remain there for a
while by making transition between them. This can be considered as a
short range attractive interaction between the two walkers.  So the
dynamics in this sector is equivalent to a problem of two
random walkers on a line having a short range attractive interaction.

We can use the same procedure of antisymmetrization to get some
more the eigenvectors.  One type of eigenvector is
obtained by antisymmetrizing about both ends simultaneously.
\be
\ket{\psi_1}= \left[ \ket{bbab\ldots abaa} - \ket{bbab\ldots
abcc}\right] - \left[ \ket{ccab\ldots abaa} - \ket{ccab\ldots
abcc}\right]
\ee

The second type of eigenvector is obtained
by antisymmetrizing two intermediate states $\ket{\ldots abcaacab\ldots}$
and $\ket{\ldots abcbbcab\ldots}$.
\be
\ket{\psi_2}= \ket{\ldots abcaacab \ldots} - \ket{\ldots abcbbcab\ldots}
\ee
Since there are $L$ positions on  the string, where $abc\beta\beta cab$
can occur, for the case of periodic boundary conditions, there are
$L$ such eigenvectors.

This can be easily generalized for the case where there are $2n$ random
walkers ($2n<L$). A typical intermediate state formed by pair wise
union of random walkers will have $n$ number of substrings
of the form $\ldots abc\alpha\alpha cab\ldots$. The eigenvector
is constructed by antisymmetrization of all these. The corresponding
eigenvalue will be $-3n$. Since these $n$ substrings can be arranged
in roughly $^LC_n$ ways, we can get so many eigenvectors.
As an example, in the case of $n=2$ one such eigenvector is given by
\be
\begin{array}{ccc}
\ket{\psi}&=&~~\ket{\ldots abcaacab\ldots abcaacab \ldots}
            -\ket{\ldots abcbbcab\ldots abcaacab \ldots} \\
          & &+\ket{\ldots abcbbcab\ldots abcbbcab \ldots}
             -\ket{\ldots abcaacab\ldots abcbbcab \ldots} \\
\end{array}
\ee
\section{Autocorrelation functions}
For the trimer model, the existence of the infinite number of
conservation laws lead to non-exponential relaxation in equilibrium
\cite{ddmb2}. It is found that the density-density autocorrelation
function decay as a power law for large times and the exponent is
different in different sectors. In most sectors the decay is
$t^{-1/4}$ but is of the form $t^{-1/2}$ in most of the sectors with
periodic irreducible strings. In some special sectors with
periodic irreducible string, stretched exponential behavior
(exp(-$\sqrt{t}$)) is found. In the null sector the autocorrelation
function is found to decay as $t^{-3/2z}$, where $z\approx 2.5$.  This
diversity of relaxational behaviour has been explained in terms of the
hard core random walkers with conserved spins (HCRWCS)
model \cite{ddmb2,ddmb3}. It has been argued on qualitative grounds that
these two models are in the same universality class. However, a strict
proof of this proposition is not yet available.

An  important assumption of the  HCRWCS
model is that the spin carried by each walker is
conserved, but does not affect the diffusive motion of the walkers
in any way. This is not strictly true in the trimer model.
However, for our model, this property can be exactly established as a
consequence of the recoloring symmetry. Thus the argument which
predict the different behaviour of
spin-spin autocorrelation function in different sectors is much
cleaner for this model than for the trimer model studied
earlier. We proceed to give the arguments which are  an adaptation of the
arguments for the trimer deposition evaporation model \cite{ddmb2,ddmb3},
in some detail.
The time dependent spin-spin autocorrelation function in the steady state
is defined by
\be
C_{\alpha \beta}^{i}(t) = \mbox{Prob($q_i(t)=\beta,q_i(0)=\alpha$)}
- \mbox{Prob($q_i(t)=\beta$)} \mbox{Prob($q_i(0)=\alpha$)}.
\ee
where Prob($x$) denote the probability of event $x$
and $q_i(t)$ is the state of the spin at the $i$th site at
time $t$. Each of
$\alpha$ and $\beta$ is either $a,b$ or $c$.  Since $\alpha$ and
$\beta$ each takes $3$ values, one can define $9$ autocorrelation
functions in this model. Among these, by time reversal symmetry
$C_{\alpha,\beta}^i(t)=C_{\beta,\alpha}^i(t)$ and also
$\sum_{\beta}C_{\alpha,\beta}^i(t)=0$. Thus there are only $3$
independent correlation functions.

As in the trimer model this autocorrelation
functions shows sector dependent decay \cite{ddmb2}.
Let us consider the  following  five different representative
sectors characterised by the irreducible strings
$1)$. $ababab\ldots$, $2).$ $acbcacbc\ldots$, $3)$. $abcabc\ldots$,
$4)$. random string and $5)$. $\phi$ (Null sector),  where $\ldots$
denotes repetition. The length of the irreducible string in the first
four cases is taken to be $L/2$.

We will first present a qualitative theory for the decay of
autocorrelation functions and then compare its predictions with our
Monte Carlo data. The behavior of the autocorrelation function can be
understood in terms of the random walk of the characters which
constitute the irreducible string.  Consider a sector labelled by an
irreducible string $\alpha_1,\alpha_2\ldots,\alpha_l$, with the length
of the irreducible string $l$ is a finite fraction of the total length
$L$.  Let $x_1,x_2\ldots,x_l$ be the positions of the characters in a
given configuration $C$ which do not get deleted under the deletion
algorithm of section $3$. Then $q_{x_i}(t)=\alpha_i$ ($i=1$ to $l$) at
all times. We can think of $x_1,x_2\ldots,x_l$ as the positions of $l$
interacting random walkers with the $n$th random walker from the left
carrying a color $\alpha_n=q_{x_n}$ with it. The positions of random
walkers, $\{x_i\}$ will change in time as $C$ changes.  The walkers
move either to the left or right but always remain on the same
sublattice. Further, they do not cross each other.

We have argued in section $4$ that the spectrum of the stochastic
matrix in this model is completely independent of the detail of the
irreducible string string sequence $\{\alpha_i\}$ and only depend up
on the length of the sequence. However, the autocorrelation involves a
weighted sum of correlations of $\{\alpha_i\}$, which gives rise to
different relaxational behavior in different sectors.

The important contribution to the autocorrelation function comes from
times when the site $i$ is occupied by character from the irreducible
string. We shall assume with out proof that the contribution coming
from times when the site $i$ is occupied by a reducible character is
qualitatively similar.

Since the hard core random walkers will remain on the  same
sublattice they were initially, we can perform a sublattice
average of $C_{\alpha,\beta}^i$. Let $\Gamma$ denotes the sublattice
which is either $A$ (odd sites) or $B$ (even sites).
In HCRWCS approximation we can write the expression for the
autocorrelation function as
\be
C_{\alpha\beta}^\Gamma(t)\cong\sum_{k=-\infty}^{+\infty}P(k,t)M^{\Gamma}(k)
\label{c_ab_2}
\ee
where $P(k,t)$ gives the joint probability of finding two particles at
a given site, one at time $0$ and the other at time $t$ such that
the difference between their labels is $2k$.  $M^{\Gamma}(k)$ is
a  measure of the correlation of the colors in the irreducible string, and
is given by
\be
M^\Gamma(k)=\sum_m \delta (q_{x_m},\alpha) \delta(q_{x_{m+2k}},\beta),
\ee
where the summation is over all the labels of the particles which
belong to the sublattice $\Gamma$. The probability $P(k,t)$ can be
obtained using the results of random walks of hard core particles
\cite{lig,alex,snmb1}.

Note that $P(k,t)$ depends only on the total length of the irreducible
string and not on its details. While $M^\Gamma(k)$ depends only on the
irreducible string and is independent of time.  This separation of the
sector dependence and time dependence is the crucial feature which
allows calculation of $C_{\alpha,\beta}^\Gamma$ in different cases.
{\it This separation is exact due to the recoloring symmetry in our
model, but only approximate in the original trimer model.}

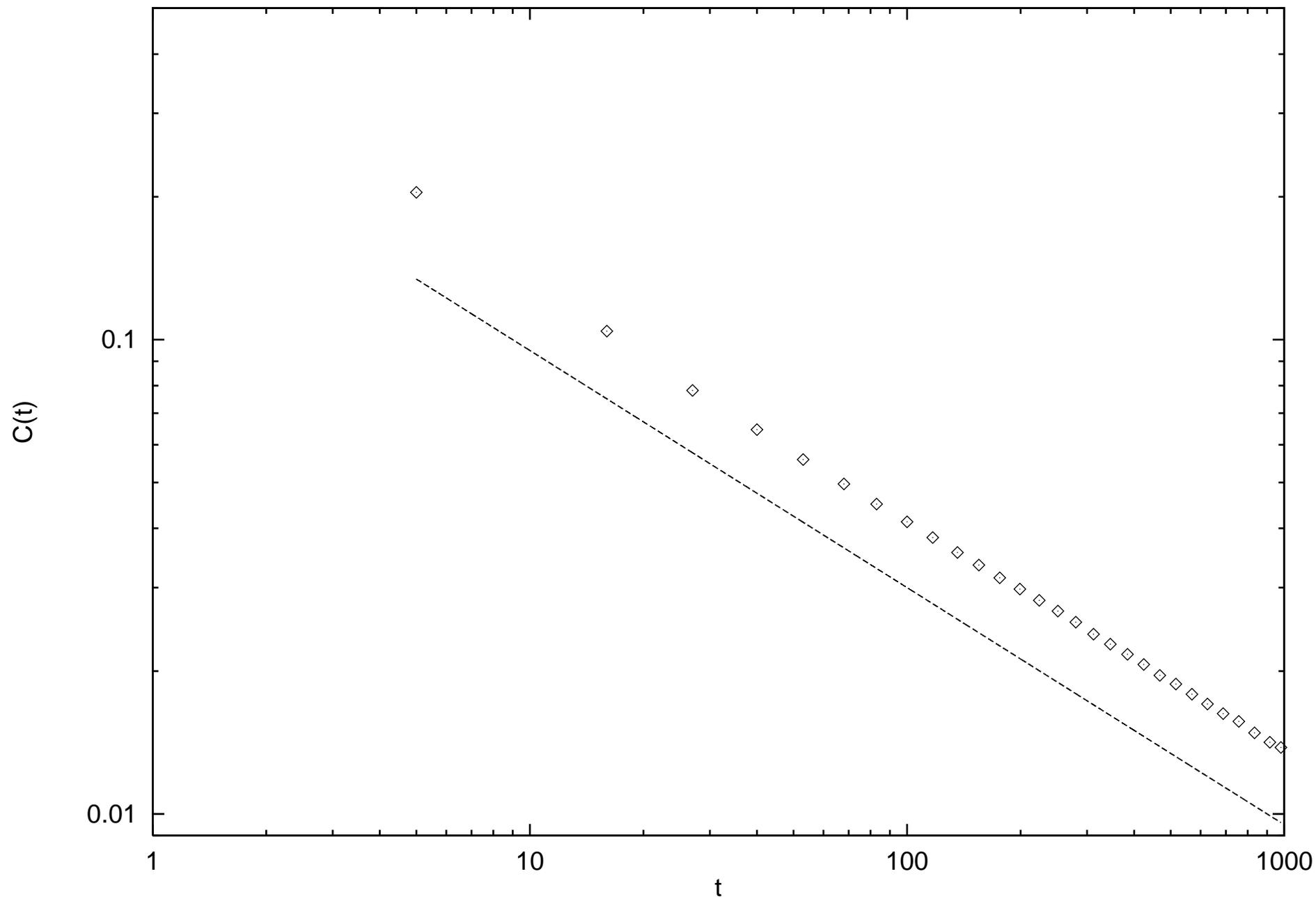
\begin{figure}
\caption{The autocorrelation function $C_{aa}^A(t)$ for the
sector $abab\ldots$ ($\diamond$). The dashed line is a plot of
$t^{-1/2}$.}
\label{abab_aa}
\end{figure}

Equation (\ref{c_ab_2}) can be used to find the behaviour of
autocorrelation function. For example consider the sector
labelled by the irreducible string $abab\ldots$.
On the $A$ sublattice $q_{x_m}=a$ for all $m$ and hence
\be
 C_{aa}^A(t)=\sum_{k=-\infty}^{+\infty}P(k,t)
\ee
This is the
density-density correlation function of hard core particles which
decays as $t^{-1/2}$ for large $t$ \cite{lig}. Our simulation gives
the same power law decay which is shown in figure \ref{abab_aa}.
{}From the symmetry between $b$ and $c$ on this sublattice and the
constraint $\sum_{\beta} C_{\alpha,\beta}^{\Gamma}(t)=0$, it can be
easily seen that $C_{\alpha,\beta}^A(t) \sim t^{-1/2}$ for all
values of $\alpha$ and $\beta$. The
behavior of the correlation functions on both the sublattices is the
same which follows from the symmetry of the irreducible string.

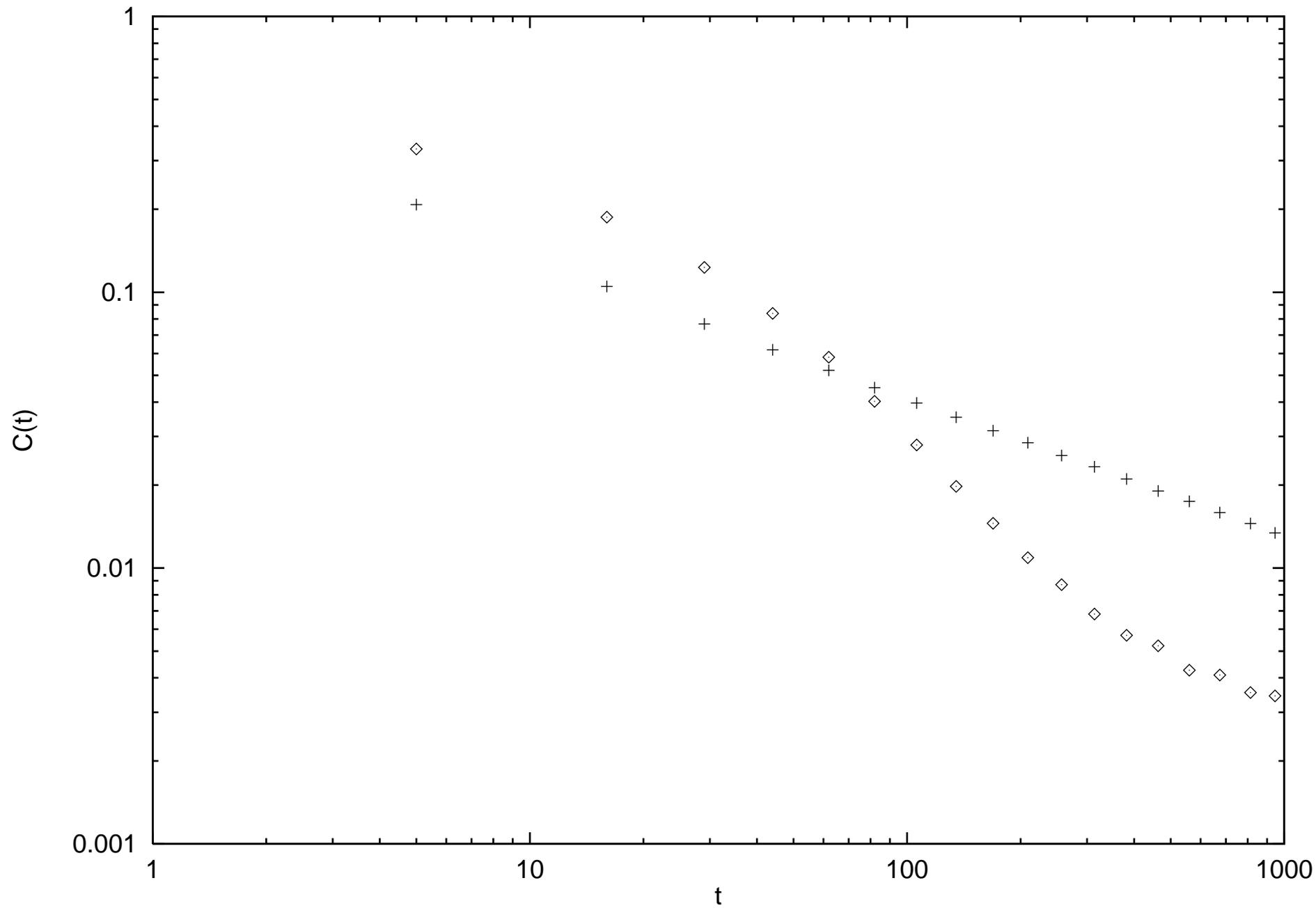
\begin{figure}
\caption{The autocorrelation functions $C_{aa}^A(t)$ ($\diamond$)
and  $C_{bb}^B(t)$ ($+$) for the sector $acbc\ldots$.  For small $t$,
$C_{aa}^A(t)$ has a stretched exponential decay of the form
exp(-$t^{-1/2}$) but at later times it crosses over to
pure diffusive decay. On the other hand $C_{aa}^B(t)$ has  pure diffusive
decay at all times.}
\label{acbc_aa}
\end{figure}

In the case of the sector labelled by the irreducible string
$acbc\ldots$, $M^A(k)$ is of the form $c_1+c_2(-1)^{k}$,
where $c_1$ and $c_2$ are constants independent of $k$.  Thus
$C_{aa}^A(t)$ has two parts. The first part decay like $t^{-1/2}$ as
explained earlier. The second part is the Fourier transform of
$P(k,t)$ which goes as exp (-$\sqrt{t}$) for large $t$.  The
stretched exponential decay dominates the short term behavior but
asymptotically the behavior will cross over to $t^{-1/2}$. By symmetry
$C_{bb}^A(t)$ should have the same decay.  On the $B$
sublattice $q_{x_m}$ is $c$ for all $m$, so the decay will be
purely diffusive ($t^{-1/2}$) as explained earlier.
These behaviours can be seen in figure \ref{acbc_aa}.

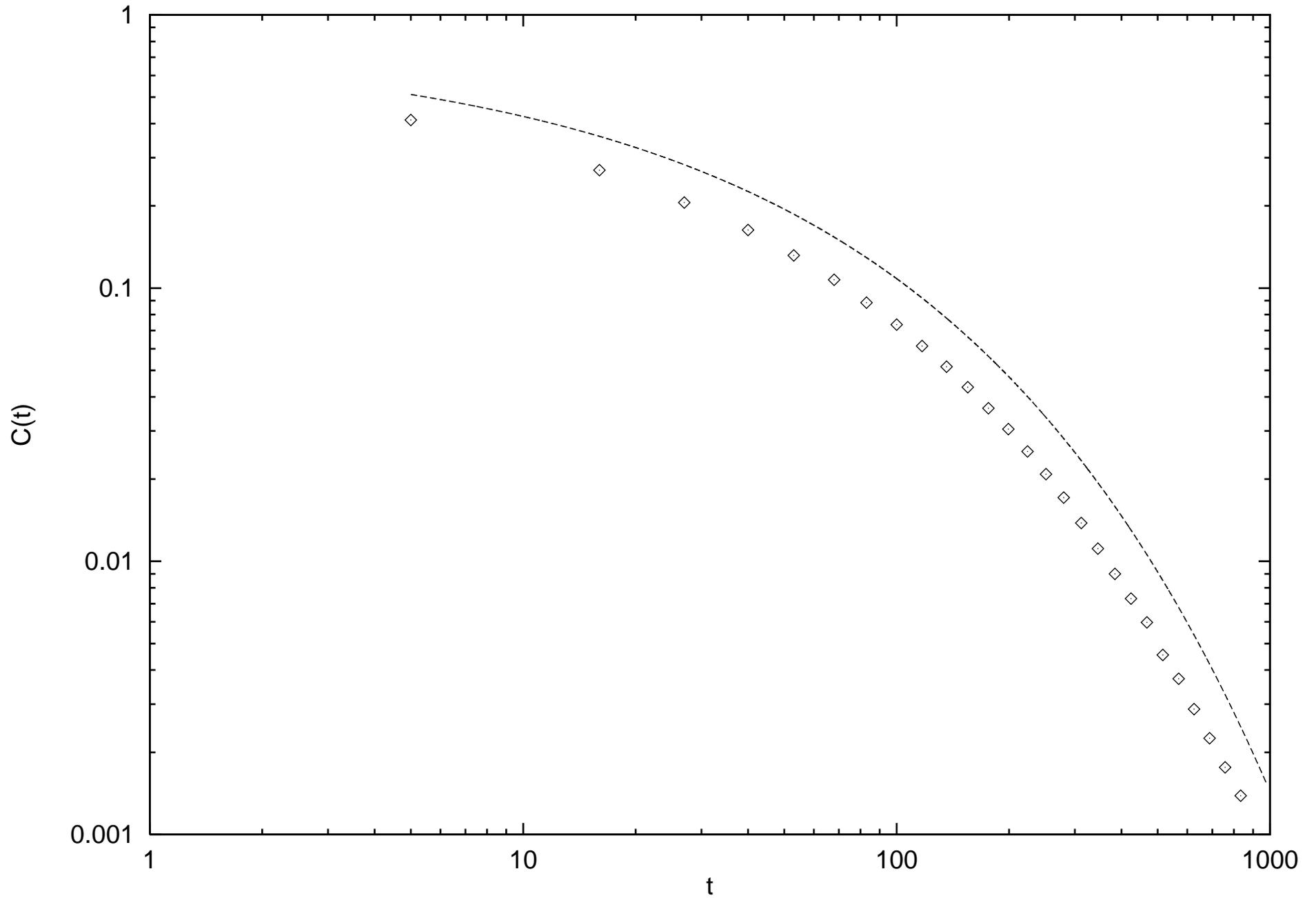
\begin{figure}
\caption{The autocorrelation function $C_{aa}^A(t)$ for the
sector $abcabc\ldots$. The decay is  stretched
exponential of the form exp(-$t^{-1/2}$) which shown by the dashed line.}
\label{abc_aa}
\end{figure}

For the sector $abcabc\ldots$, the decay of $C_{aa}^A(t)$ is shown if
figure \ref{abc_aa}. The leading behavior of the correlation function
can be shown to be stretched exponential decay by a similar argument
as in the previous case.  By symmetry of the irreducible string the
behaviour should be the same for all the correlation functions and on
both the sublattices.  This has been confirmed in our simulation.

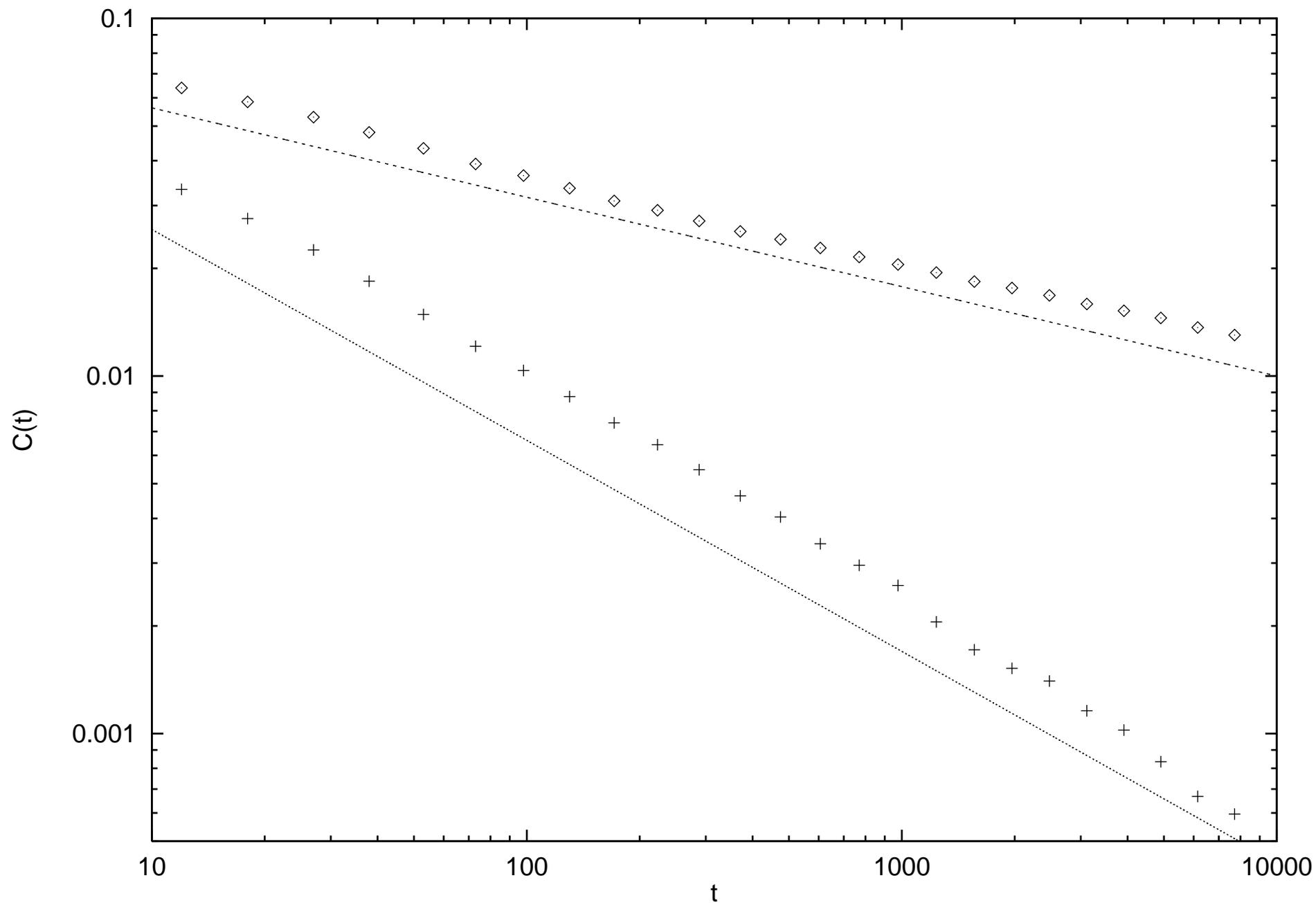
\begin{figure}
\caption{The decay of $C_{aa}^A(t)$ ($\diamond$)in the sector where
irreducible string is random,  and in the null sector ($+$).
These decays are approximately given by $t^{-1/4}$ (dashed line),
and  $t^{-0.59}$ (dotted line).}
\label{null_aa}
\end{figure}

When the irreducible string is random, $M^{\Gamma}(k)$ is significant
only for small values of $k$. For small $k$, $P(k,t)\sim t^{-1/4}$ for
large $t$ \cite{ddmb2}.  Therefore the autocorrelation has a
$t^{-1/4}$ decay which can be seen in figure \ref{null_aa}.

Figure \ref{null_aa} also shows the decay of $C_{aa}^A(t)$ in the null
sector.  It is a power law decay with the value of the exponent
$\approx 0.59$.  The same power law decay found for other values of
$\alpha$ and $\beta$ and on both sublattices, which follows from
symmetry.  Barma {\it et.al.} have obtained same estimate for this
exponent for the trimer model in the null sector \cite{ddmb2}.  This
is an evidence that the dynamics in these two models for the case of
the initial condition in which all sites are in the same state,
belongs to the same universality class. We can understand the
behaviour of the correlation functions in sectors where the
irreducible string has a finite density using the HCRWCS model, this
approach is no longer useful when studying the null sector.  An
analytical understanding of the temporal decay of autocorrelation
functions in this sector is still lacking.

\section{Spatial Correlation Function in the
Steady State}

Let $C_{\alpha,\beta}(r)$ be the probability that, in the steady
state, spins at two sites which are separated by a distance $r$ are
$\alpha$ and $\beta$.  We calculate this equal time spatial
correlation function for the null sector.  Let us assume periodic
boundary conditions.  Let $S$ be the string corresponding to a given
configuration on this lattice with sites $i$ and $i+r$ are occupied by
$\alpha$ and $\beta$.  Let $S_1$ be the substring of length $r-1$
formed by spins between $\alpha$ and $\beta$ and $S_2$ be the
substring of length $L-r+1$ formed by the remaining spins .  The
endpoints of the substring $S_2$ are $\alpha$ and $\beta$.  Let
$IS(S_1)$ and $IS(S_2)$ be the irreducible strings corresponding to
$S_1$ and $S_2$.  Since $IS(S)=\phi$, $IS(S_2)$ will be $IS(S_1)$
reflected about one end.  We may write $IS(S_2)\equiv IS(S_1)^{-1}$.
\be
C_{\alpha,\beta}(r)=\frac{\sum_{IS}D(IS,r-1)D_{\alpha,\beta}(IS^{-1},L-r+1)}
{D(\phi,L)}
\label{corr1}
\ee
where the summation is over all the distinct irreducible strings of
$S_1$. $D(IS,r-1)$ is the size of the sector labeled by
$IS$ on a lattice of length $r-1$ with free boundary
conditions and $D_{\alpha,\beta}(IS^{-1},L-r+1)$ is the size of the
sector labeled by  $IS^{-1}$ on a lattice of length
$L-r+1$ with boundary sites fixed at $\alpha$ and $\beta$.

Though we can calculate exact expressions for each of the above
quantities, we derive the asymptotic behaviour by taking $L
\rightarrow \infty, r<<L$. The length of the irreducible string
$l\le r$, therefor in this limit one can make the approximation
$D_{\alpha,\beta}(IS,L-r+1)\approx D_{\alpha,\beta}(\phi,L-r-l+1)$.
In the case of free boundary condition the size of a
sector depends only on the length of the irreducible string
(\ref{sec_size1},\ref{sec_size2}) and we may replace the sum over
distinct irreducible strings by a sum over the lengths of the
irreducible strings as follows.
\be
C_{\alpha,\beta}(r)=\frac{\sum_lN(l)D(l,r-1)D_{\alpha,\beta}(\phi,L-r-l+1)}
{D(\phi,L)} \label{corr1.1}
\ee
Where $N(l)$ is the number of distinct
irreducible strings of length $l$, which is given by $3\times2^{l-1}$. For
large $L$, $D_{\alpha,\beta}(\phi,L-r-l+1)\sim (2
\sqrt{2})^{L-r-l+1}/(L-r-l+1)^{3/2}$.  Hence in the summation over $l$
only terms corresponding to small $l$ will contribute. We can find the
asymptotic $r$ dependence by approximating the sum by only the $l=0$
term. Then we have
\be C_{\alpha,\beta}(r)\sim
D(\phi,r-1)D_{\alpha,\beta}(\phi,L-r+1)/D(\phi,L).
\ee For large $L$
gives
\be C_{\alpha,\beta}(r)\sim\frac{f(\alpha,\beta)}{r^{3/2}}
\ee
for large
r, where $f(\alpha,\beta)$ is a constant independent of $r$ but
depends on $\alpha$ and $\beta$. From the symmetry between different
colors, it follows that $f(\alpha,\beta)=k~(1-3\delta_{\alpha,\beta})$
\section{Numerical Diagonalization of the Stochastic Matrix
in the Null Sector.}  As mentioned in the section $6$, the decay of
autocorrelation function in the null sector suggests that the dynamics
in this sector belongs to a new universality class. The description of
dynamics in terms of the random walk of the characters of the
irreducible string is not possible in this sector. One way of studying
this sector is to directly diagonalize the stochastic matrix for small
sizes. However this task is not very easy because the size of the
matrix to be diagonalized grows very fast with the size of the lattice
($\sim (2\sqrt{2})^L/L^{3/2}$).  One can make use of various
symmetries of the model to block diagonalize the matrix first and
hence reduce the size of the matrix to be diagonalized.  By using
translation, reflection and the recoloring symmetry described in
section $4$ we have been able to diagonalize the stochastic matrix for
lattice sizes up to $20$, in the fully symmetric subspace.  The
algorithm used here for diagonalizing the matrix using the various
symmetries is similar to the one we have used for studying trimer
model \cite{pthd1}. For details of the algorithm we refer the reader
to this paper.

In table $1$, we have listed the size of the matrix ($N_{R,T,C}$)
obtained after using all the three symmetries and the value of the
largest nonzero eigenvalue in the symmetric subspace in the null
sector for lattice sizes $L$ from $2$ to $20$.  For comparison we also
list the total size ($N$) of the null sector, and also the reduced
matrix size ($N_C$) if only the recoloring symmetry C is taken into
account.  We see that the additional recoloring symmetry gives a
significant reduction in the size of the matrix to be diagonalized.

\begin{table}
\begin{tabular}{|r|r|r|r|r|r|} \hline
 $L$&$\hfill N\hfill$ &$\hfill N_{C}\hfill$& $N_{R,T,C}$&$\hfill
 \lambda_L \hfill$&$\hfill z_L\hfill$ \\
\hline
            2  &       3 &      1 &    1 &           &            \\
\hline
            4  &      15 &      3 &    2 & -10.00000  &            \\
\hline
            6  &      87 &     12 &    4 & -3.108548  &  2.881701  \\
\hline
            8  &     543 &     55 &   10 & -1.629235  &  2.245691  \\
\hline
           10  &    3543 &    271 &   26 & -0.974243  &  2.304369  \\
\hline
           12  &   23823 &   1399 &   93 & -0.632500  &  2.369330  \\
\hline
           14  &  163719 &   7470 &  338 & -0.438272  &  2.379750  \\
\hline
           16  & 1143999 &  40931 & 1474 & -0.318921  &  2.380690  \\
\hline
           18  & 8099511 & 228918 & 6801 & -0.240934  &  2.380807  \\
\hline
           20  &57959535 & 1301778&33746 & -0.187477  &  2.381040  \\
\hline
\end{tabular}
\label{table1}
\end{table}

The largest eigenvalue of the stochastic matrix is $0$.
The difference between  the
largest and the second largest eigenvalues gives the gap in the
spectrum of the stochastic matrix. We have computed this gap for
lattice sizes from $2$ to $20$. By using the finite size scaling
relation $\Delta_L \sim
L^{-z}$, the  dynamical exponent $z_{L}$ for a lattice of size $L$
is estimated from the relation
\be
z_{L} = \log(\frac{\lambda_{L}}{\lambda_{L-2}})/\log(\frac{L-2}{L}).
\ee
The estimate of $z_L$ for various values of $L$ are shown in table $1$.
The convergent of $z_L$ seemed to be quite good, and the
estimate of the extrapolated value of $z$ corresponding to $L=\infty$
is $2.39 \pm 0.05$. The error bar reflects our subjective
estimate of the possible systematic errors in the extrapolation.
This value is in good agreement
with the value of $z=2.5\pm0.15$ for the trimer model in the null
sector \cite{pthd1}, and is evidence that both  these  models are
in the same  universality  class.

We thank M. Barma for introducing us to the model, and  for his
continued interest and discussions throughout the  course of
this work. We have benefited from several discussions with
Peter. B. Thomas in the earlier stages of this project.

\newpage

\end{document}